\documentclass[twoside,11pt,preprint]{article}

\usepackage{jmlr2e}

\usepackage[utf8]{inputenc}
\usepackage{subcaption}
\usepackage{amsmath}

\usepackage{lastpage}
\jmlrheading{23}{2023}{1-\pageref{LastPage}}{5/12; Revised NA}{NA}{21-0000}{Ali Ismail-Fawaz, Angus Dempster, Chang Wei Tan, Matthieu Herrmann, Lynn Miller, Daniel Schmidt, Stefano Berretti, Jonathan Weber, Maxime Devanne, Germain Forestier, and Geoffrey I. Webb}

\ShortHeadings{Multi-Comparison Matrix}{Ismail-Fawaz et al.}
\firstpageno{1}

\begin{document}

\title{An Approach to Multiple Comparison Benchmark Evaluations that is Stable Under Manipulation of the Comparate Set}

\author{%
    \name Ali Ismail-Fawaz$^1$ \email ali-el-hadi.ismail-fawaz@uha.fr\\
    \name Angus Dempster$^2$ \email Angus.Dempster1@monash.edu\\
    \name Chang Wei Tan$^2$ \email Chang.Tan@monash.edu\\
    \name Matthieu Herrmann$^2$ \email matthieu.herrmann@monash.edu\\
    \name Lynn Miller$^2$ \email lynn.miller1@monash.edu\\
    \name Daniel F.~Schmidt$^2$ \email Daniel.Schmidt@monash.edu\\
    \name Stefano Berretti$^3$ \email stefano.berretti@unifi.it\\
    \name Jonathan Weber$^1$ \email jonathan.weber@uha.fr\\
    \name Maxime Devanne$^1$ \email maxime.devanne@uha.fr\\
    \name Germain Forestier$^{1,2}$ \email germain.forestier@uha.fr\\
    \name Geoffrey I.~Webb$^2$ \email geoff.webb@monash.edu\\
    $^1$ IRIMAS, Université Haute-Alsasce\\
    $^2$ Department of Data Science and Artificial Intelligence, Monash University\\
    $^3$  Department of Information Engineering, University of Florence (UNIFI)
}

\editor{My editor}

\maketitle

\begin{abstract}%
    The measurement of progress using benchmarks evaluations is ubiquitous in computer science and machine learning.  However, common approaches to analyzing and presenting the results of benchmark comparisons of multiple algorithms over multiple datasets, such as the critical difference diagram introduced by Dem\v{s}ar (2006), have important shortcomings and, we show, are open to both inadvertent and intentional manipulation. To address these issues, we propose a new approach to presenting the results of benchmark comparisons, the Multiple Comparison Matrix (MCM), that prioritizes pairwise comparisons and precludes the means of manipulating experimental results in existing approaches. MCM can be used to show the results of an all-pairs comparison, or to show the results of a comparison between one or more selected algorithms and the state of the art. MCM is implemented in Python and is publicly available.
\end{abstract}


\section{Introduction}

Many areas of computer science stimulate and measure progress on performance benchmarks.  Many of these benchmarks involve multiple tasks. For example, the influential UCI Machine Learning Repository~\citep{dua2023uci}, contains numerous machine learning tasks. When a new machine learning method is proposed, in order to understand its place relative to the current state-of-the-art, it is accepted practice to compare its performance to that of current methods on a large set of relevant UCI tasks~\citep{demvsar2006statistical}.

We call the methods being compared in such a study \emph{comparates}.  When $m$ comparates are assessed on $n$ tasks, the evaluation results in an outcome (e.g., classification error) for each comparate on each task, a total of $m \times n$ outcomes, which frequently is in the order of thousands. It is difficult to make sense of these raw results. In consequence, a number of methods have been developed to produce comprehensible summaries~\citep{demvsar2006statistical, garcia2008extension}. However, these have the serious deficiency that the conclusions they support about the relative performance of two comparates can be altered (inadvertently or otherwise)---indeed, can change fundamentally---when other comparates are added or removed from a comparison. As a result, they are relatively straightforward to game should a researcher wish to change the impression given by the results. Further, they are based on null hypothesis significance testing (NHST), an approach that is under increasing attack~\citep{benavoli2017time, berrar2022pvalues}. 

The Critical Difference diagram (CD diagram), introduced by \citet{demvsar2006statistical}, is widely used to show the relative performance of different comparates over multiple tasks. This method of presenting results has found widespread adoption. In the field of time series classification, it is now ubiquitous \citep[e.g.,][]{rocket,hc2}. On the face of it, the CD diagram provides both \emph{groupwise} and \emph{pairwise} comparisons. The former place the performance of each comparate within the context of all other comparates while the latter directly compare two specific comparates. Groupwise comparisons are achieved by ordering all comparates on the mean of the ranks of relative performance on each task. Pairwise comparisons are based on a `critical difference' in this mean rank or, in more recent variants~\citep{benavoli2016should}, the Wilcoxon signed-rank test (hereafter \emph{Wilcoxon test})~\citep{wilcoxon1992individual} with multiple test correction. However, we show that this approach does not provide a stable pairwise comparison, with the method being open to gaming.

Subsequent work~\citep{benavoli2017time} has concentrated on improving the procedure for establishing whether or not the pairwise difference in accuracy between comparates is statistically significant, including various Bayesian approaches \citep{benavoli2015bayesian,corani2017statistical,yu2017new,berrar2022pvalues,jansen2022statistical,wainer2022bayesian}. However, the use of the CD Diagram remains largely unchanged since \citet{demvsar2006statistical}, despite well-known shortcomings relating to the underlying statistical methods.  Bayesian approaches do not appear to have gained widespread adoption so far.

The CD diagram uses mean rank to establish a global ordering of comparates over all tasks. This presents at least three major difficulties. First, rank ignores the magnitude of differences between comparates, and so can both hide large differences and exaggerate insignificant differences. Second, rank is sensitive to which comparates are included in the comparison (e.g., the relative rank of two comparates, $c_i$ and $c_j$, can be affected by whether comparates $c_u$ and/or $c_v$ are also included in the comparison)~\citep{benavoli2016should, berrar2022pvalues}. Third, per the so-called ``no free lunch'' theorem \citep{wolpert1997lunch}, as an aggregate measure of performance over multiple datasets, the rank of all algorithms will converge as the number of datasets increases~\citep{berrar2022pvalues}.

We argue, therefore, that there should be less emphasis on a global aggregate measure of performance, such as mean rank, and more emphasis on the pairwise differences between comparates and the relative performance of different comparates on different tasks (which is likely to be of the greatest practical significance).  Furthermore, we argue that these comparisons should be stable, that is, our conclusions when comparing two comparates should not be affected by the addition or exclusion of other comparates to a study.

To this end, we propose a new approach to summarizing multiple comparison benchmarks, the Multiple Comparison Matrix (MCM), and provide it as a community resource in the form of open source software. The core idea of the new proposal is to clearly separate groupwise and pairwise comparisons. 
Further, to address the limitations of traditional statistical hypothesis testing, the proposed approach emphasizes descriptive over inferential statistics.

In particular, the MCM:
\begin{itemize}
    \item  emphasizes pairwise comparisons;
    \item  provides easy-to-digest overviews of relative performance on a benchmark of multiple tasks; 
    \item ensures that pairwise outcomes between any two comparates $c_i$ and $c_j$:
    \begin{itemize}
\item will be constant from one study to another on the same benchmark; and
        \item  cannot be gamed by manipulating the set of other comparates included in a study.
    \end{itemize}
\end{itemize}

The rest of the paper is organized as follows. In Section~\ref{sec:background}, we present an overview of the task of multiple comparison benchmark evaluation, summarize existing methods, and discuss serious deficiencies therewith. In Section~\ref{sec:MCM}, we lay out the case for and describe our MCM method. Finally, in Section~\ref{sec:conclusion}, we provide concluding remarks.

\section{Background and Current Benchmarking Methods}\label{sec:background}

We examine the problem of summarizing the outcomes of evaluation of $m$ comparates $\mathcal{C}=\{c_1,\ldots,c_m\}$ on multiple tasks $\mathcal{T}=\{t_1,\ldots,t_n\}$ with respect to a single measure of performance $\gamma: \mathcal{C}, \mathcal{T} \rightarrow\mathbb{R}$ that assesses the performance of a comparate $c\in\mathcal{C}$ on a task $t\in\mathcal{T}$. We use time series classification as a running example, where each comparate is a time series classification algorithm. Each task is to learn a classifier $\lambda$ from a time series training set, and the measure of performance is the accuracy (strictly speaking, zero-one loss) of $\lambda$ on the given time series test set.  Time series classifiers form the set of comparates $\mathcal{C}$, classification on 108 datasets from the UCR archive~\citep{ucrArchive} the tasks $\mathcal{T}$, and classification accuracy the performance measure $\gamma$.

In all cases, we use previously published results. As different algorithms have been assessed on different subsets of the 128 datasets in the archive, we use only those datasets for which results are available for all our comparates. The excluded datatsets are: 15 datasets that have variable length and/or have missing values; one dataset (\emph{Fungi}) that has only one sample for at least one class label in the training set; and the following four datasets, which are not included in the published results for some classifiers used in this study: \emph{HandOutlines, FordB, NonInvasiveFetalECGThorax1} and \emph{NonInvasiveFetalECGThorax2}.

\sloppy We illustrate multiple comparisons using as comparates different sets of times series classification algorithms selected from the following 24 for which we had ready access to the relevant performance outcomes: 
Arsenal~\citep{hc2};  
Bag-Of-Symbolic Fourier approximation-Symbols (BOSS)~\citep{boss}; 
CAnonical Time-series CHaracteristics 22 (Catch22)~\citep{catch22}; 
Contractable BOSS (C-BOSS)~\citep{cboss}; 
Canonical Interval Forest (CIF)~\citep{cif}; 
Diverse Representation Canonical Interval Forest (DrCIF)~\citep{hc2}; 
Hierarchical Vote Collective of Transformation-based Ensembles (HIVE-COTE 1.0 referred to as HC1)~\citep{hc1,hc1-extension}; 
HIVE-COTE 2.0 (HC2)~\citep{hc2}; HYbrid Dictionary–Rocket Architecture (HYDRA)~\citep{hydra}; 
Hybrid-InceptionTime (H-InceptionTime)~\citep{ismail-fawaz2022hccf}; 
InceptionTime~\citep{inception}; 
MiniROCKET \citep{minirocket}; Multiple pooling operators ROCKET 
(MultiROCKET)~\citep{multirocket}; 
ProximityForest~\citep{proximityforest}; 
Residual Network (ResNet)~\citep{ismail2019deep,fcn_resnet};  
Random Interval Spectral Ensemble (RISE)~\citep{hc1-extension}; 
RandOm Convolutional KErnel Transform (ROCKET)~\citep{rocket}; 
Spatial BOSS (S-BOSS)~\citep{sboss}; 
Shapelet Transform Classifier (STC)~\citep{stc}; 
Supervised Time Series Forest (STSF)~\citep{stsf}; 
Temporal Dictionary Ensemble (TDE)~\citep{tde}; 
Time Series Combination of Heterogeneous and Integrated Embedding Forest (TS-CHIEF)~\citep{tschief}; 
Time Series Forest (TSF)~\citep{tsf}; and 
Word ExtrAction for time SEries cLassification (WEASEL)~\citep{weasel}.

\subsection{Ranking Comparates}\label{sec:ranks}

The CD diagram~\citep{demvsar2006statistical} is the primary current method used for multi-comparate multi-task benchmarking.
This diagram presents (1) a groupwise comparison using the mean rank of each comparate; and (2) a pairwise comparison showing which pairs of comparates are and are not significantly different in their performance: see the example in Figure \ref{fig:cd-diagram}.

Mean rank is calculated as follows. Each comparate $c_i$ is assigned a rank $R_{c_i}^t$ on each task $t$ based on relative score on the measure of performance $\gamma$:
\begin{equation}
    R_{c_i}^t=1 + \left|\{c_j\in\mathcal{C}\setminus c_i: \gamma(c_j,t)\succ\gamma(c_i,t)\}\right| + \tfrac{1}{2} \cdot \left|\{c_j\in\mathcal{C}\setminus c_i: \gamma(c_i,t)=\gamma(c_j,t)\}\right|,
\end{equation}

where $\succ$ means \emph{better than}. For example, a higher value of accuracy is \emph{better than} a lower value, while a lower value of error is \emph{better than} a higher value.
Each comparate is then assigned an Average Rank (AR) by averaging its ranks over all $n$ tasks in $\mathcal{T}$,
\begin{equation}
    \mathrm{AR}^{\mathcal{T}}_{c_i}=\sum_{t\in\mathcal{T}}R_{c_i}^t/n.
\end{equation}

The position of the comparates on the diagram is based on their AR. The lower the AR, the better the assessment of aggregate performance relative to the competitor comparates.  (A Friedman test~\citep{friedman1940comparison} is performed to assess whether there exists a significant groupwise difference in ranks.)  Critically, however, the rank of a given comparate on a given task, $R_{c_i}^t$, and therefore the average rank of a given comparate over all tasks, $\mathrm{AR}^{\mathcal{T}}_{c_i}$, depends on the performance of all of the other comparates, $\mathcal{C} \setminus c_i$, on each task.

\subsection{Pairwise Comparisons with the CD Diagram}

A number of approaches have been developed for evaluating the statistical significance of pairwise differences between comparates, including the Nemenyi test, and the Wilcoxon signed-rank test.

\subsubsection{Nemenyi Test}

\citet{demvsar2006statistical} initially proposed the Nemenyi test~\citep{nemenyi1963test} as a post hoc test to determine the statistical significance of  pairwise differences between comparates.  The Nemenyi test defines a `critical difference' in terms of mean rank: the difference in performance between two comparates is considered statistically significant if the difference between their mean ranks is larger than this critical difference.

However, this test has a number of limitations \citep{benavoli2016should}. The test only considers rank, and thus considers only the number of tasks on which a comparate achieves better or worse performance than other comparates, ignoring the magnitude of those differences. In other words, it treats marginal differences in performance in exactly the same way as  substantial differences.  A further issue is that the test is unstable with respect to the set of comparates included in the assessment. Adding or removing a single comparate can change the pairwise conclusions drawn between the remaining comparates \citep{benavoli2016should}.

\subsubsection{Wilcoxon Signed Rank Test}

\citet{demvsar2006statistical} proposed that the Wilcoxon test~\citep{wilcoxon1992individual} should be used as the statistical test in cases where there are only two comparates (with the Nemenyi test used as the pairwise test when there are multiple comparates, as above).

\citet{benavoli2016should} argued that the Wilcoxon test should be used for all pairwise comparisons, replacing the Nemenyi test, to avoid the issues outlined above, and the Wilcoxon test is now increasingly used in place of the Nemenyi test for pairwise comparisons \citep[e.g.,][]{rocket,hc2}.

The Wilcoxon test is used to determine the statistical significance of the differences in performance between two comparates.  The Wilcoxon test uses only the outcomes (e.g., classification accuracy) for the given pair of comparates, and is unaffected by the outcomes for any other comparates.  The Wilcoxon test also takes into account, albeit in an indirect and ordinal way, the magnitude of the differences in performance between comparates \citep{demvsar2006statistical,benavoli2016should}.

The resulting $p$ value represents the probability of observing the given differences in performance (represented by the test statistic) between a pair of comparates if the null hypothesis---that the median difference in performance between the pair of comparates is zero---is true \citep{demvsar2006statistical,berrar2022pvalues}.  In other words, a `small' $p$ value suggests that the observed differences in performance are improbable if the median difference in performance between two comparates was zero.

When used in conjunction with the CD diagram, the Wilcoxon test is applied to all pairs of comparates in $\mathcal{C}$, typically with a correction to control familywise error, such as the Holm correction \citep{demvsar2006statistical,garcia2008extension,benavoli2016should}.

\subsection{Limitations of the CD Diagram}\label{sec:limitations-cd-diagram}

\begin{figure}
    \centering
    \includegraphics[width=0.95\linewidth]{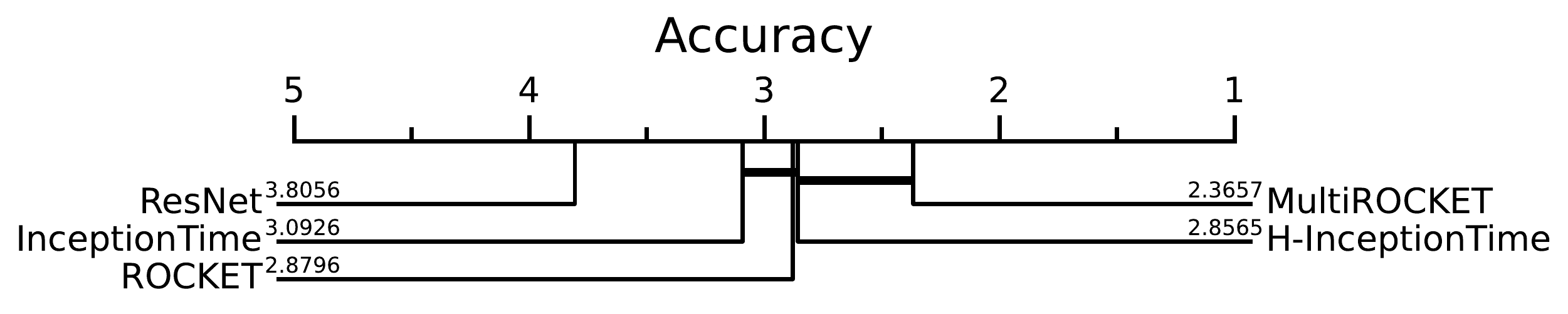}
    \caption{An example CD diagram. The best average rank is placed to the right and the worst to the left.}
    \label{fig:cd-diagram}
\end{figure}

Figure~\ref{fig:cd-diagram} shows an example CD diagram, comparing the performance in terms of classification accuracy of five time series classifiers on 108 datasets from the UCR archive~\citep{ucrArchive}.  Note that the comparates are ordered from right to left in terms of mean rank (i.e., smaller mean ranks indicating higher classification accuracy appear on the right).  Classifiers where the pairwise differences in classificaion accuracy are \textit{not} statisticaly significant, per a Wilxocon test with Holm correction, are joined with a black line.
In this example, the diagram shows that MultiROCKET is the `best' classifier in terms of average rank among this set of comparates on this set of tasks.
Note, however, that lines join MultiROCKET and H-InceptionTime, as well as H-InceptionTime, InceptionTime, and ROCKET, suggesting that the differences in classification accuracy between these pairs of comparates are not statistically significant.

A major strength of the CD diagram is that it provides a simple summary of a large amount of information that is easy to understand and digest. However, this simplicity carries with it a number of limitations.  In what follows, we argue that there are three principal limitations of the CD diagram, namely: \emph{(i)} instability of the average rank, \emph{(ii)} insufficient attention to the magnitude of differences in performance, and \emph{(iii)} undesirable consequences of using multiple testing corrections.

\subsubsection{Instability of the Average Rank}

\begin{figure}
    \centering
    \includegraphics[width=0.85\linewidth]{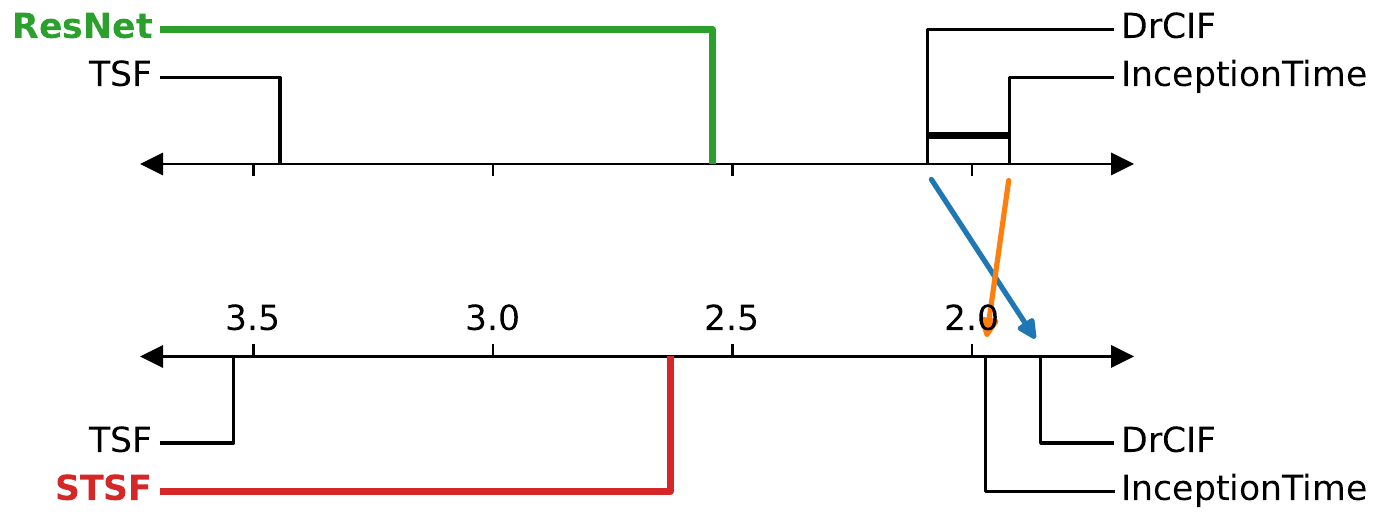}
    \caption{Manipulation of the ranks of DrCIF and InceptionTime---and the statistical significance of their pairwise differences---by inclusion of similar comparates. When ResNet is replaced by STSF, DrCIF moves from a `worse' to a `better' rank, and the pairwise differences between DrCIF and InceptionTime change from being not statistically significant to statistically significant.}
    \label{fig-swap-1}
\end{figure}

\begin{figure}
    \centering
    \includegraphics[width=0.85\linewidth]{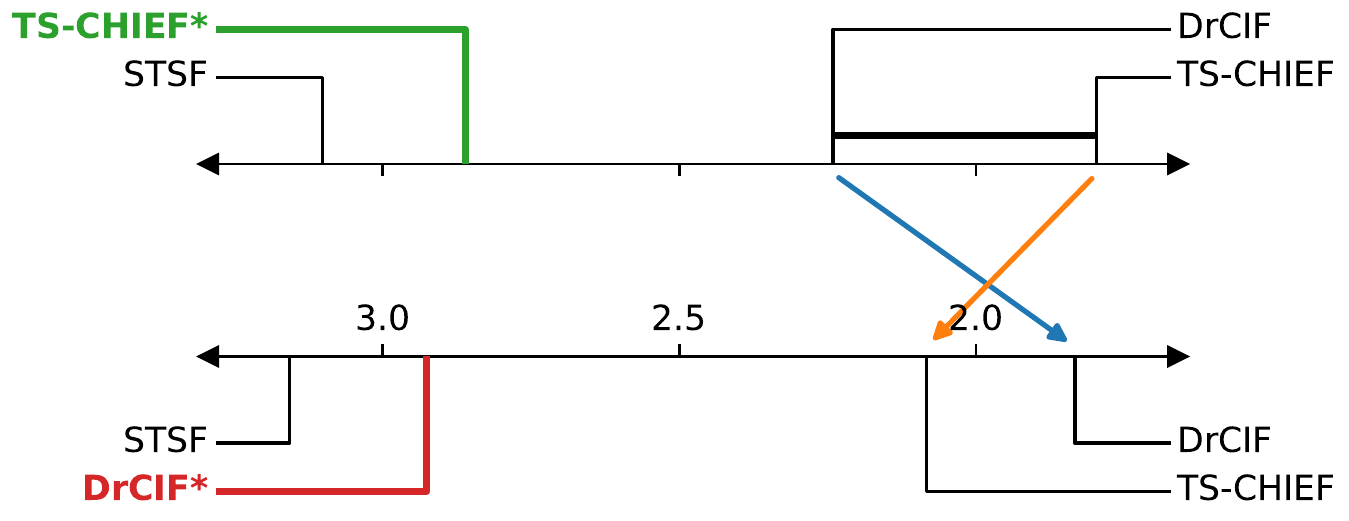}
    \caption{Manipulation of the ranks of DrCIF and TS-CHIEF---and the statisical significance of their pairwise differences---by inclusion of weakened comparates. When a weakened variant of TS-CHIEF is replaced by a weakened variant of DrCIF, DrCIF moves from a `worse' to a `better' rank and the pairwise differences between DrCIF and TS-CHIEF change from being not statistically significant to statistically significant.}
    \label{fig-swap-2}
\end{figure}

The CD diagram orders comparates by mean rank (per Section \ref{sec:ranks}, above).  However, mean rank can change with the addition or removal of one or more comparates---that is, the relative order of some set of comparates, $C$, can change with the addition or removal of one or more other comparates---as shown in Figures \ref{fig-swap-1} and \ref{fig-swap-2}.  (As in Figure \ref{fig:cd-diagram}, two comparates for which the pairwise differences are \textit{not} statistically significant per a Wilcoxon tests with Holm correction are joined with a black line.  For simplicity, in these examples, statistical significance is only shown for those comparates which are ``swapping'' ranks.)  In particular, Figure \ref{fig-swap-1} shows that by removing ResNet from and adding STSF to the set of comparates, DrCIF ``swaps'' ranks with InceptionTime---i.e., moves ahead of InceptionTime in terms of mean rank---and the pairwise differences between DrCIF and InceptionTime change from not being statistically significant to being statistically significant.  (In this example, ResNet is a `weaker' deep learning algorithm than InceptionTime and STSF is a `weaker' interval method than DrCIF.)

Figure \ref{fig-swap-2} demonstrates how the addition of a weaker variant of a given comparate (e.g., a different version of a given classifier using fewer parameters or different hyperparameter tuning) can be used to manipulate mean rank and pairwise statistical significance.  In this example, we simulate weaker variants of each of TS-CHIEF and DrCIF (denoted by an asterisk) by using a weighted average of the accuracies for each comparate with the accuracies of a weaker comparate (Catch22).  The inclusion of a weaker variant of a given comparate can significantly elevate the rank of the original comparate and change the pattern of the statistical significance of pairwise differences between comparates.  The original comparate will be more accurate than its weaker variant (by definition) on a large number of tasks, improving the mean rank of the original comparate, and resulting in a small $p$ value under the Wilcoxon signed-rank test (which can, in turn, ``push'' pairwise differences between the original comparate and other comparates outside the threshold for statistical significance under the Holm correction).

For this reason, we propose that comparates should be ordered on a statistical measure that is independent of the addition or removal of other comparates. This way, the relative order of different comparates will remain stable from study to study, and cannot be manipulated (inadvertently or deliberately).

\subsubsection{Insufficient Attention to the Magnitude of Wins and Losses}\label{sec:magnitude-rank}

Mean rank counts the number of tasks for which a comparate `wins' or `loses' (e.g., achieves higher or lower classification accuracy) relative to each other comparate in a study. It pays no attention to the magnitude of the differences in performance. A new comparate can achieve a small average rank through many small wins, while also suffering large losses. Given a choice between comparates $c_i$ and $c_j$, where $c_i$ has a 90\% chance of an insubstantial loss to $c_j$ and a 10\% chance of a major gain, many would prefer to use $c_i$, but mean rank strongly favors $c_j$.

The Wilcoxon test has some sensitivity to the magnitudes of wins, but this does little to remedy the situation when the primary comparison is with respect to rank, as in the CD diagram.

\subsubsection{Null Hypothesis Significance Testing}

\paragraph{Criticisms of NHST}

The use of statistical significance testing to compare performance on benchmark tasks is coming under increasing questioning. \citet{benavoli2017time} argue that  NHST, including the Wilcoxon test, are not ideally suited to multiple-comparate and multiple-task benchmarking for four reasons \citep[see also][]{berrar2022pvalues}.

\textbf{First}, NHST \emph{does not estimate the probability of the hypothesis of interest} (the \emph{alternative hypothesis}, or in other words, the hypothesis that there is a difference between comparates given the observed outcomes). Instead, NHST produces the probability that the observed outcomes $O$ (or more extreme) would be observed if the null hypothesis $H_0$ were true, $p(O\mid H_0)$, i.e., that there is no difference between comparates.
Benavoli~et~al.~argue that a more useful assessment is $p(H_0\mid O)$.

\textbf{Second}, they point out that the null hypothesis can always be rejected just by adding a few more examples, a point that we return to in Section~\ref{sec:mult-test}.

\textbf{Third}, they argue that the NHST does not provide any information about the magnitude of differences between comparates, even when the $p$ value is extremely small.
For instance, consider a case where one comparate has a $10^{-4}$ higher accuracy than another on most tasks in a set of benchmark tasks.
The $p$ value (e.g., from a Wilcoxon test) in this case may be very small, potentially giving the false impression that the magnitude of the differences in accuracy between the classifiers is substantial.
The $p$ value fails to capture, however, any information about the magnitude of the difference in performance between comparates. Note that this problem is related to, but separate from, the issue raised in Section~\ref{sec:magnitude-rank} in relation to ranks failing to capture the magnitude of difference in performance between comparates.

\textbf{Finally}, as the $p$ value represents the probability of observing the given outcomes (conditional on the null hypothesis), rather than the null hypothesis itself, a large $p$ value (on the basis of which we might conclude, informally, that the differences are not statistically significant) does not provide evidence as to whether or not the null hypothesis itself is true~\citep{lecoutre2022significance}.

\paragraph{Inferential vs Descriptive Statistics}

The $p$ value derived from a statistical hypothesis test is often used as an inferential (rather than descriptive) statistic. A key difference between descriptive statistics and inferential statistics is that the former describe precisely, in some sense, the empirical statistical properties of the results, while the latter attempt to make inferences about the population from which the experiments were generated. That is, in practical terms, inferential statistics attempt to quantify or provide predictions about how a comparate would perform on new or unseen data randomly drawn from the same sources. In this sense, inferential statistics make stronger statements about the results of the comparisons, but only do this by making stronger assumptions. 

\sloppy Consider the case of applying the Wilcoxon test to two comparates $c_i$ and $c_j$ on $n$ tasks $\mathcal{T}=t_1,\ldots, t_n$ resulting in performance measures $\gamma(c_i,t_1),\ldots,\gamma(c_i,t_n)$ and $\gamma(c_j,t_1),\ldots,\gamma(c_j,t_n)$. The (two--tail) Wilcoxon test returns the probability $p$ that $\gamma(c_i,t_1)-\gamma(c_j,t_1),\ldots,\gamma(c_i,t_n)-\gamma(c_j,t_n)$ (or more extreme) would be observed if those values were an iid sample from a distribution $\Omega$ that is symmetric around zero. Used as an inferential statistic, the null hypothesis that the distribution from which these observations were drawn is symmetric is rejected if $p\leq\alpha$, where $\alpha$ is the chosen significance level. Recall that the interpretation of this $p$ value is the likelihood of seeing a test statistic as large, or larger, than the one observed, just by chance, if we selected our comparates `ahead of time', without access to the data.

In classical scientific studies this paradigm is workable because experiments are usually conducted by collecting fresh data. However, the reality is that in many benchmark situations, researchers choose their comparates by selecting algorithms and hyperparameters that perform well on the given benchmark. They do not collect `new' benchmark data by sampling again from the problems which define the benchmark. As such, it is difficult to conceive from what meaningful distribution the resulting performance scores could be considered an iid sample and, consequently, how the corresponding test statistic and $p$ value can be used as inferential statistics.

Neverthelesss, these test statistics and $p$ values are generally useful measures of divergence between two or more bodies of data points (for example, the classification accuracies of two comparates) and, as such, provide a quantitative {\em descriptive} measure of the differences in performance between comparates.

\subsubsection{The Use of Multiple Test Corrections}\label{sec:mult-test}

It is widely accepted that the chosen significance level (that is, a threshold value below which a given $p$ value will be considered statistically significant, typically $0.05$) should be adjusted when performing multiple tests, e.g., Wilcoxon tests between all pairs of comparates \citep[for example]{demvsar2006statistical,benavoli2016should,berrar2022pvalues}.

Multiple test corrections seek to control the risk that any null hypothesis should be rejected in error when many null hypotheses are tested at once. The Holm correction~\citep{holm1979simple} is commonly used for this purpose. Given significance level $\alpha$, the Holm correction ensures that the risk is no more that $\alpha$ that any of a family of null hypotheses will be rejected in error.
Under the Holm correction, the $p$ values are arranged in an ascending order.
This correction then adapts the significance level applied to the $i^{th}$ $p$ value from $\alpha$ to $\frac{\alpha}{n+1-i}$, and stops as soon as the smallest $p$ value is larger than the adjusted significance level (that is, the null hypothesis is not rejected for this $p$ value and all larger $p$ values), with $n$ being the total number of $p$ values, i.e., the total number of comparisons.

However, we argue that it is more important to control the risk with respect to each pair of comparates rather than with respect to a study as a whole, and we show that the use of multiple testing corrections introduces highly undesirable side effects that allow benchmark results to be manipulated, and undermine the attempt to present a consistent, stable, comparison of multiple comparates over multiple tasks.

The key problem is that the choice of the number of comparates in a given study, and/or the particular comparates used in a given study, affect the potential for one comparate to be found to outperform another \citep{benavoli2017time,kruschke2018bayesian}. This allows for the statistical significance of the differences between given comparates to be manipulated (inadvertently or otherwise) by adding or removing other comparates to the comparison.

\begin{figure}[ht]
    \begin{subfigure}{0.5\linewidth}
        \centering
        \includegraphics[width=0.995\linewidth]{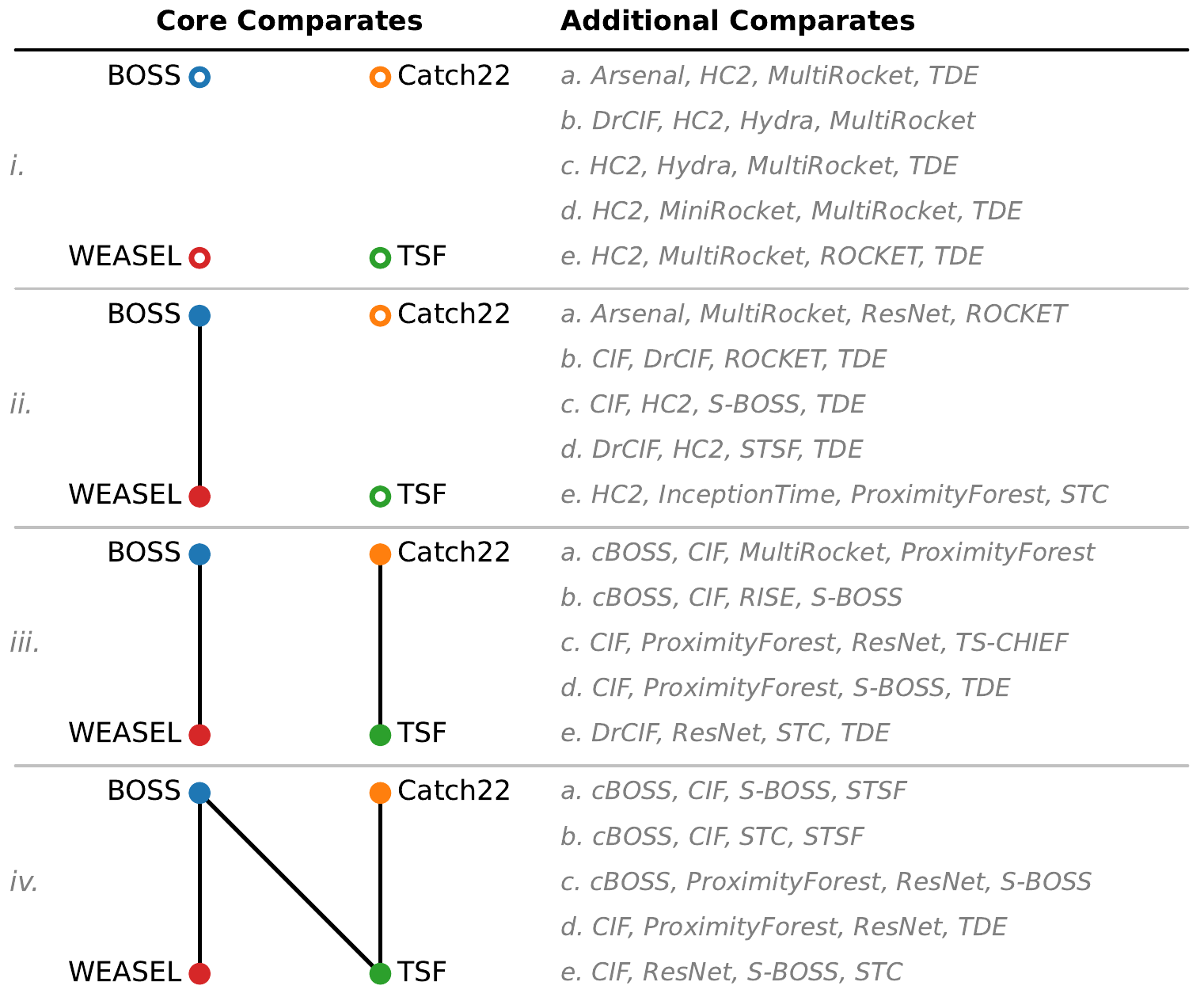}
        \caption{\null}
    \end{subfigure}
    \begin{subfigure}{0.5\linewidth}
        \centering
        \includegraphics[width=0.995\linewidth]{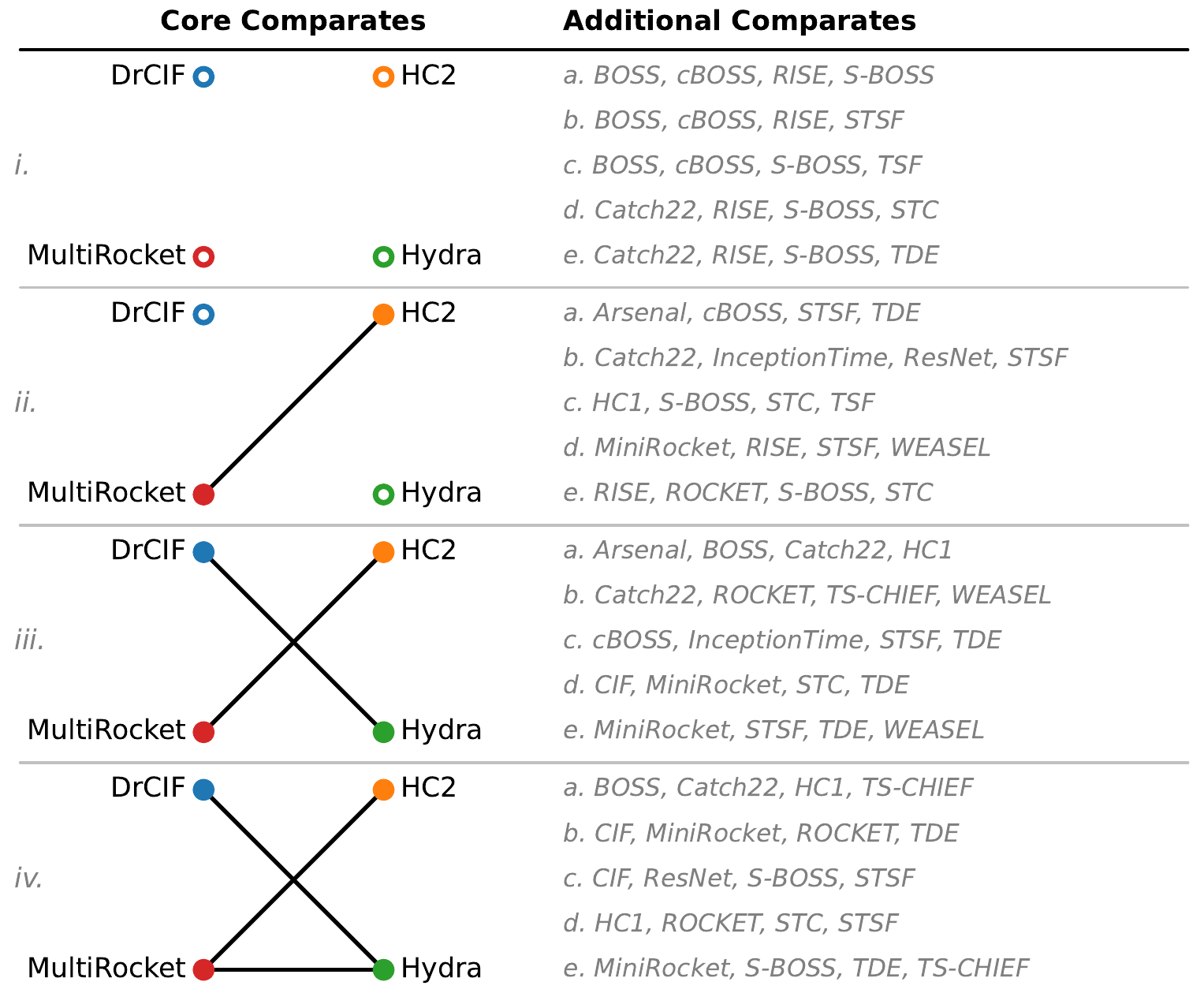}
        \caption{\null}
    \end{subfigure}
    \caption{Two examples of the instability of pairwise significance under the Holm correction: (a) for comparates BOSS, Catch22, TSF, and WEASEL; and (b) for comparates DrCIF, HC2, Hydra, and MultiRocket. The statistical significance of pairwise differences between comparates is affected by which additional comparates are included in the comparison. In each example, four different patterns of statistically-significant pairwise differences---(i), (ii), (iii), and (iv)---are shown in the left column (pairs for which the differences are \textit{not} statistically significant per the Wilcoxon signed-rank test with Holm correction are connected with a black line), and randomly-selected examples of additional comparates, which produce the given pattern of statistically-significant pairwise differences are shown in the right column.}
    \label{fig-pairwise}
\end{figure}

To demonstrate the instability of the statistical significance of pairwise differences between comparates when using the Holm correction, we show the impact on statistical significance of changes in the set of comparates. 
Two examples of such instability---i.e., where the statistical significance of pairwise differences changes depending on the particular set of comparates included in the comparison---are shown in Figure~\ref{fig-pairwise}.  In each example, we first select a ``core'' set of four comparates, and then repeatedly combine this core set with different sets of four additional comparates.  For each combination of core and additional comparates, we perform the Wilcoxon test for all pairs with the Holm correction, noting in each case which pairwise differences within the core set are statistically significant.  Lines connecting core comparates indicate that the pairwise differences are \textit{not} statistically significant.

Figure~\ref{fig-pairwise} shows results for core sets consisting of (a) BOSS, Catch22, TSF, and WEASEL; and (b) DrCIF, HC2, Hydra, and MultiRocket.  (These examples are drawn from results for 23 different comparates over 108 datasets from the UCR archive.)  For example, as shown in Figure~\ref{fig-pairwise} (a), combining comparates BOSS, Catch22, TSF, and WEASEL with any of the five sets of comparates listed on the right for pattern (ii)---e.g., Arsenal, MultiRocket, ResNet, and Rocket---produces the pattern of statistical significance of pairwise differences shown on the left, i.e., the pairwise differences between BOSS and WEASEL are \textit{not} statistically significant.  However, combining the same core comparates with any of the sets of additional comparates listed in (i), (iii), or (iv) produces a different pattern of statistical significance.

The examples in Figure~\ref{fig-pairwise} show that it is possible to ``game'' the statistical significance of pairwise differences between comparates, inadvertently or otherwise, simply by adding or removing other comparates.  In~many cases, many different sets of additional comparates---tens, hundreds, or even thousands, depending on the size of the set of possible comparates---produce the same pattern of statistical significance of pairwise differences.  For example, in Figure~\ref{fig-pairwise}(b), there are $123$ different sets of additional comparates, which produce pattern (i), $1{,}876$, which produce pattern (ii), $680$, which produce pattern (iii), and $1{,}197$, which produce pattern (iv).  For simplicity, only five randomly-selected combinations of additional comparates are shown for each pattern of statistical significance in Figure~\ref{fig-pairwise}.

This issue arises where a multiple testing correction, such as the Holm correction, makes the threshold for statistical significance depend on the $p$ values for all pairs of comparates.  That is, whether or not the pairwise differences for a given pair of comparates depends on the $p$ values for all other pairs of comparates.  Therefore, the addition or removal of comparates with small $p$ values can ``shift'' pairwise differences above or below the threshold for statistical significance.  For a given pair of comparates, the addition or removal of one or more other comparates can make an otherwise statistically-significant difference insignificant, and vice versa.

A further issue is that multiple test corrections control the risk of any algorithm being found to outperform another when it only does so by chance. However, it does this at the expense of increasing the risk of mistakenly regarding as chance findings, results correctly showing that an algorithm outperform others. It is not clear why the research community should give primacy to one of these risks over another. It is very attractive to be able to claim that a new algorithm is not significantly less powerful than the current state-of-the-art, and the use of a multiple test correction means that the proponent of a new algorithm need only add sufficient algorithms to a comparison to obtain such a finding.

An important point to make here is that to avoid the ``data dredging'' problem, the multiple testing correction adjustment should actually be based not on the number of comparates that were finally decided on, and compared in the paper, but should also include the totality of variations of an algorithm that the researchers tried, and discarded, while developing their new comparate. Again, this is not really feasible, pointing to a conclusion that attempting to do so at all is probably fruitless.

A further problem is that different studies can find different outcomes for pairwise comparisons of the same two competitors when the number of comparates differ between studies. A multiple test correction for a study with fewer comparates might reject the null hypothesis and find a significant difference between comparates $c_i$ and $c_j$, while a study with more comparates fails to reject the null hypothesis and thus finds no significant difference between $c_i$ and $c_j$ on the basis of exactly the same evidence.
We argue that this is not a sound basis on which to produce an evidence base for a discipline.

\subsection{An Alternative Approach}

As noted earlier, recent work has attempted to address some of these issues, particularly in relation to improved statistical significance testing for pairwise differences between comparates, e.g., \citep{benavoli2015bayesian}.  We highlight, in particular, the approach proposed by \citet{benavoli2017time}, noting that one of the coauthors, Janez Dem\v{s}ar, originally proposed the CD diagram.

\citet{benavoli2017time} argued that the Wilcoxon test, or any similar test, should be abandoned and proposed in their place a new Bayesian test modeled on the Wilcoxon test. For a comparison between two comparates $c_i$ and $c_j$, the Bayesian signed rank test produces a probability distribution over $c_i$ being meaningfully better than $c_j$, $c_j$ being meaningfully better than $c_i$, and $c_i$ not being meaningfully different from $c_j$.
Let $\textbf{z}=[z_0,z_1,...,z_i,...z_q]$, assuming it follows a Dirichlet Process (DP), denote the vector of difference in performance between $c_i$ and $c_j$ on $q$ tasks including a pseudo observation, a hyper parameter of the DP, $z_0$. The final mathematical formulation of the above detailed distribution is the following:
\begin{equation}
    \begin{split}
        \theta_l &= \sum_{i=0}^q\sum_{j=0}^q \omega_i \omega_j \mathbf{I}_{(-\infty,-2r)}(z_i+z_j)\\
        \theta_e &= \sum_{i=0}^q\sum_{j=0}^q \omega_i \omega_j \mathbf{I}_{(-2r,2r)}(z_i+z_j)\\
        \theta_e &= \sum_{i=0}^q\sum_{j=0}^q \omega_i \omega_j \mathbf{I}_{(2r,\infty)}(z_i+z_j),
    \end{split}
\end{equation}

where $\mathbf{I}_A(x) = 1$ if $x \in A$, the weights $\omega_i$ follow a Dirichlet distribution $D(s,1,1,...,1)$ and $r$ is the ``rope'' value that sets the interval of which the two classifiers have no significant difference.
Given that the distribution $\theta_l, \theta_e \,\&\, \theta_r$ does not have a closed form solution, the probability distribution is generated using a Monte Carlo sampling on the weights $\omega_i$.

\begin{figure}
    \centering
    \vspace{-4em}
    \includegraphics[width=0.65\linewidth]{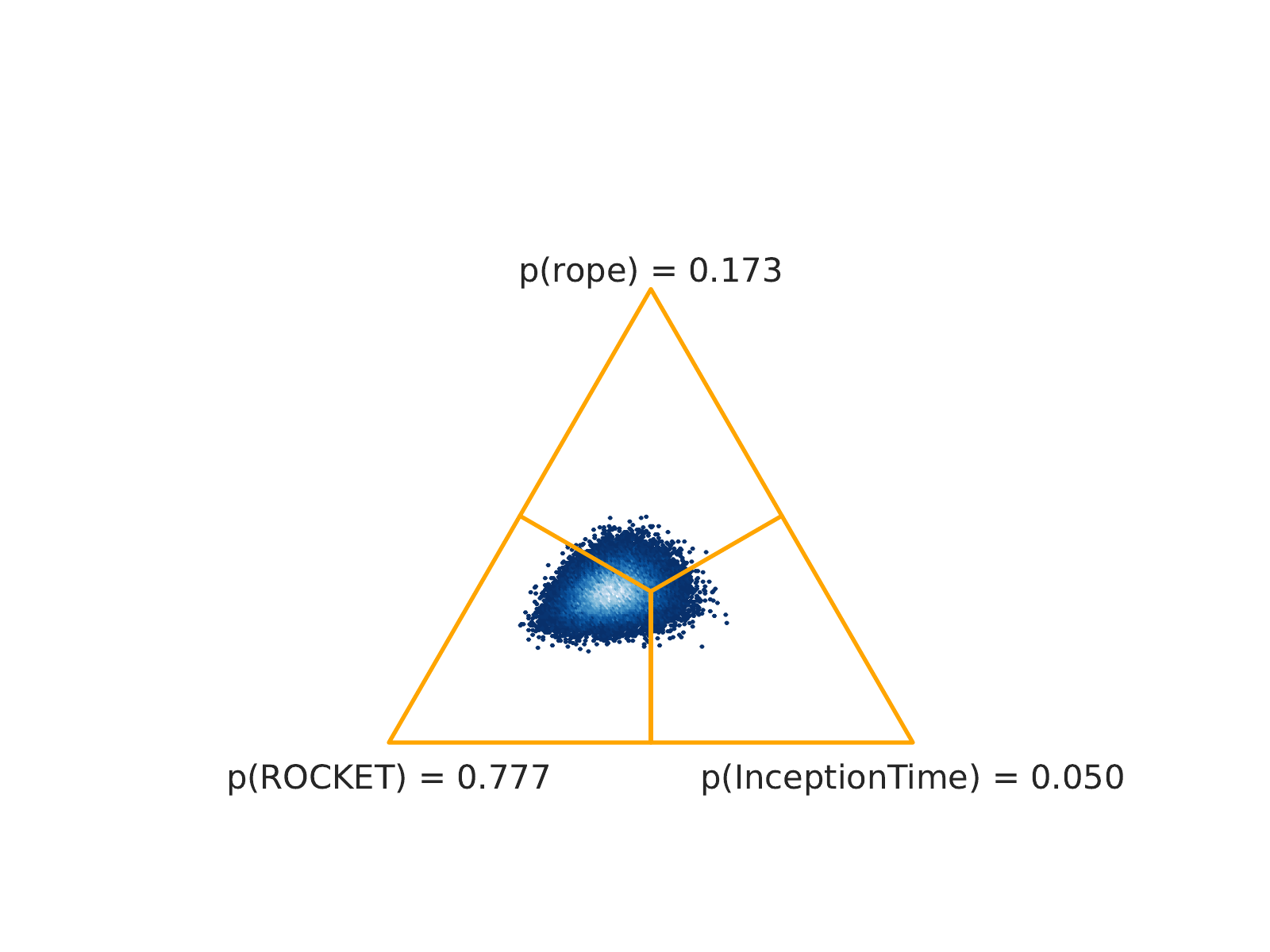}
    \vspace{-2em}
    \caption{A visualization of the Bayesian Signed Rank Test proposed in~\citet{benavoli2015bayesian} as a replacement for the Wilcoxon Signed Rank Test. As illustrated, the Bayesian test provides information on the probability that the null hypothesis would be true given the outcome classification metric of both InceptionTime and ROCKET on the 108 tasks of the UCR archive.}
    \label{fig:bayesian_triangle}
\end{figure}

An example of this Bayesian test is presented in Figure~\ref{fig:bayesian_triangle} comparing InceptionTime~\citep{inception} and ROCKET~\citep{rocket} on 108 datasets from the UCR archive~\citep{ucrArchive}.
The triangle shows that the probability that ROCKET would beat InceptionTime is $77.7\%$ with only $17.3\%$ probability that both classifiers are not meaningfully different. 

The approach proposed by \citet{benavoli2017time} focuses on the means of measuring the statistical significance of pairwise differences between classifiers, as an alternative to the Wilcoxon signed-rank test, and is not mutually exclusive with the present work.  We see the present work as complementary to efforts to improve the soundness of the statistical testing of differences between comparates.  As an alternative to the Wilcoxon signed-rank test, the probabilities computed via the Bayesian signed-rank test could be used in place of the $p$ values derived from the Wilcoson signed-rank test in the MCM.

\section{The Multi-Comparison Matrix}\label{sec:MCM}

We seek to develop techniques for evaluating $m$ comparates $\mathcal{C}$ on multiple tasks $\mathcal{T}$ with respect to a single measure of performance $\gamma$ that:

\begin{itemize}
    \item give primacy to pairwise comparisons between comparates;
    \item emphasize descriptive statistics over statistical hypothesis testing;
    \item give pairwise comparisons $\delta(c_i,c_j)$ between any two comparates $c_i\in\mathcal{C}$ and $c_j\in\mathcal{C}$ such that $\delta(c_i,c_j)$ is invariant to $\mathcal{C}\backslash\{c_i,c_j\}$ (i.e., no pairwise comparison will change with addition or deletion of other comparates and will not change from study to study); 
    \item order comparates so as the relative order of any two comparates $c_i\in\mathcal{C}$ and $c_j\in\mathcal{C}$ is invariant to $\mathcal{C}\backslash\{c_i,c_j\}$ (i.e., the order of $c_i$ and $c_j$ will not change with the addition or deletion of other comparates and will not change from study to study); and
    \item provide a good tradeoff between the amount of information presented and the informativeness of that information.
\end{itemize}

To this end, we propose to provide a grid of pairwise comparison statistics: see the examples in Figures \ref{fig:mcm-example}, \ref{fig:mcm-example-row-not-in-col}, and \ref{fig:mcm-example-row-in-col}.  We provide an open source implementation at \url{https://github.com/MSD-IRIMAS/Multi\_Comparison\_Matrix}. The proposed Multi-Comparison Matrix (MCM) preserves the pairwise comparison between each pair of comparates $c_i$ and $c_j$, by default ordering the comparates based on the average performance measure $\gamma$.
Each cell of this matrix includes three pairwise statistics between $c_i$, the comparate for the row, and $c_j$, the comparate for the column.
These three statistics are:

\begin{itemize}
    \item The mean of $\gamma(c_i,t)-\gamma(c_j,t)$ over all $t\in\mathcal{T}$.
    \item A Win Tie Loss count for $c_i$ over all the tasks in $\mathcal{T}$.
    \item A $p$ value for a Wilcoxon Signed Rank Test.
\end{itemize}

Note that despite concerns about the use of statistical significance testing for benchmarking, we retain the use of the Wilcoxon test, and the associated $p$ value.  While we believe that there is a case for excluding formal statistical hypothesis tests, we understand that this would be a step too far for many, and hence we give primacy to descriptive statistics.  However, following from the issues discussed above, we encourage consideration of the $p$ as a \textit{descriptive} statistic, that is, as a measure of the ``strength'' of the difference between comparates, but not as an inferential statistic---that is, not as suggesting likelihood or probability of a similar or equivalent difference in accuracy between a given pair of comparates on new or unseen data (that is, ``out of benchmark'').

Nonetheless, given the long tradition of statistical significance testing in benchmarking, we allow the option of specifying a significance level, by default $0.05$, and highlight in bold the statistics in cells for which the $p$ value falls below this threshold. 

Our Python implementation of this method takes as input a pandas\footnote{https://pandas.pydata.org/} dataframe of the performance measure $\gamma$, with a row for each comparate $c_i\in\mathcal{C}$ and a column for each task $t\in\mathcal{T}$.

By default the MCM is generated with all comparates present in both the rows and the columns, resulting in $m \times (m-1) / 2$ comparisons. Alternatively, separate lists of comparates for the rows and columns can be specified, $\mathcal{C}_{\mathrm{row}}$ and $\mathcal{C}_{\mathrm{col}}$. In this case there are $|\mathcal{C}_{\mathrm{row}}|\times|\mathcal{C}_{\mathrm{col}}|-|\mathcal{C}_{\mathrm{row}}\cap\mathcal{C}_{\mathrm{col}}|$ comparisons.

\subsection{Full Pairwise Multi-Comparison Matrix}\label{sec:heatmap}

\begin{figure}
    \centering
    \includegraphics[width=1.0\linewidth]{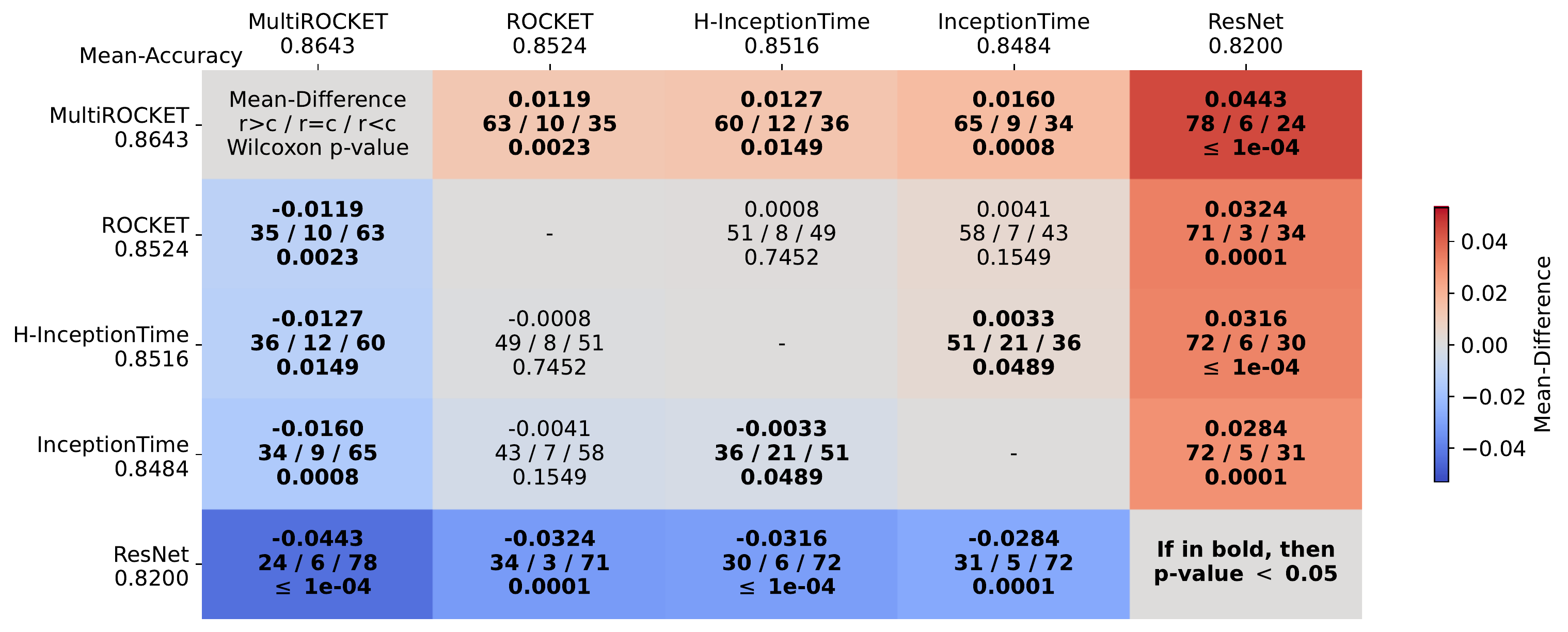}
    \caption{MCM showing all pairwise comparisons between MultiROCKET, ROCKET, H-InceptionTime, InceptionTime, and ResNet on the 108 datasets of the UCR archive. In this setup, the full pairwise comparison is presented.}
    \label{fig:mcm-example}
\end{figure}

By default, all pairwise comparisons between comparates are presented.  The average performance measure $\gamma$ (e.g., classification accuracy) over all tasks $\mathcal{T}$ is shown next to the comparate labels. An example showing the results for five comparates over 108 datasets from the UCR archive~\citep{ucrArchive} is illustrated in Figure~\ref{fig:mcm-example}. The colors of the Heat Map represent the values of the mean difference in $\gamma$. If this difference is positive (in red) this means that the comparate in the row wins by more on average (in value) than the comparate in the column.  (For example, in Figure \ref{fig:mcm-example}, the top right cell is red, reflecting that MultiROCKET (row) is on average more accurate than ResNet (column). Conversely, if the value is negative (in blue), then the comparate in the column wins by more on average (in value) than the comparate in the row. The text of each cell is represented in \textbf{BOLD} if the $p$ value is lower than a given threshold, in this case: $0.05$.

This format of the MCM is useful when presenting comparisons for a benchmark review such as in~\citet{ismail2019deep}. In such a review, information about all the pairwise comparisons is needed in order to show the strengths and weaknesses of each comparate. For instance, this format of the MCM can be used to show when a comparate that performs poorly compared to the winning comparate may perform well on some datasets compared to other SOTA comparates.

\subsection{Focused Pairwise Multi-Comparison Matrix}

If a study focuses on a few specific comparates, for example, when a new algorithm is proposed, it is often useful to focus on the comparison of those few against many existing alternatives. When using the MCM in these situations, it can be formatted to show just the results needed to compare proposed comparates to the existing SOTA. This allows the reader to focus on the relevant results. The proposed comparates are listed in one of the rows or columns of the matrix, while the SOTA approaches are listed in the other. Both rows and columns are ordered by the performance measure $\gamma$. The information in each cell is the same as that used in the full pairwise comparison and is presented from the perspective of the row comparate, showing how it performs compared to each of the included column approaches.

\begin{figure}
    \centering
    \includegraphics[width=0.75\linewidth]{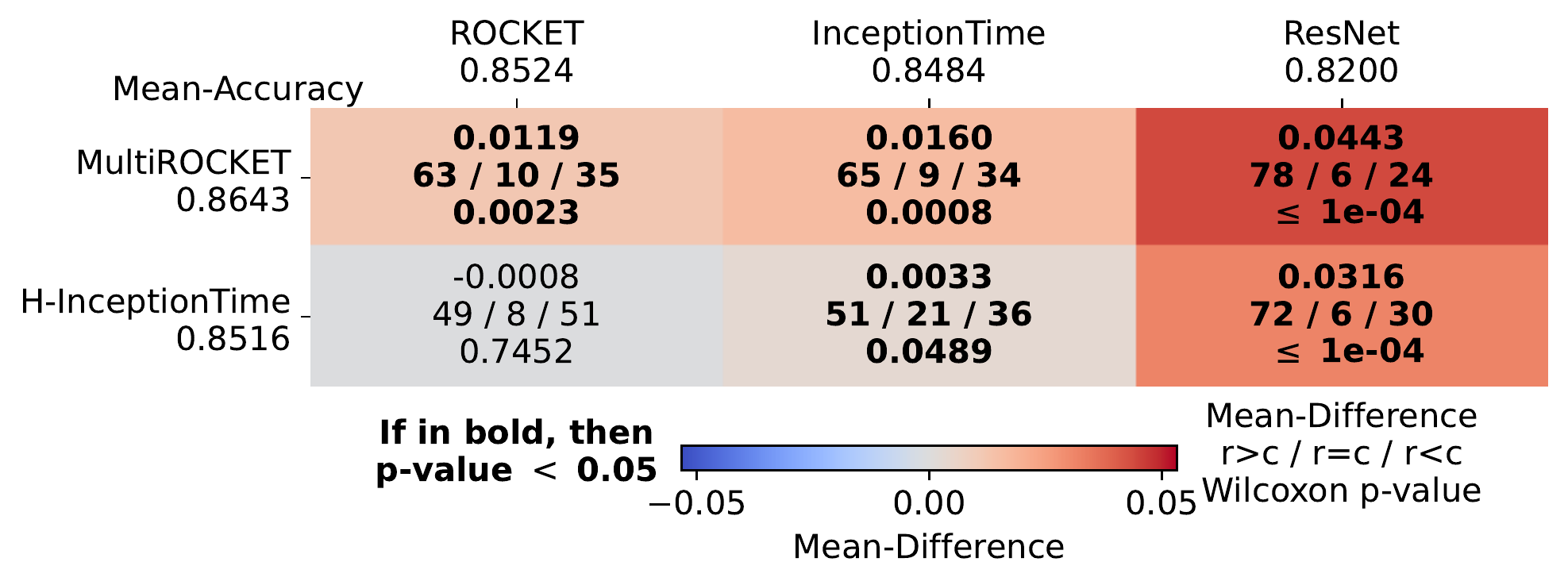}
    \caption{MCM showing pairwise comparisons between MultiROCKET and H-InceptionTime  and each of ROCKET, InceptionTime, and ResNet on the 108 datasets of the UCR archive. In this setup, the row and column comparates are mutually exclusive.}
   \label{fig:mcm-example-row-not-in-col}
\end{figure}

\begin{figure}
    \centering
    \includegraphics[width=0.9\linewidth]{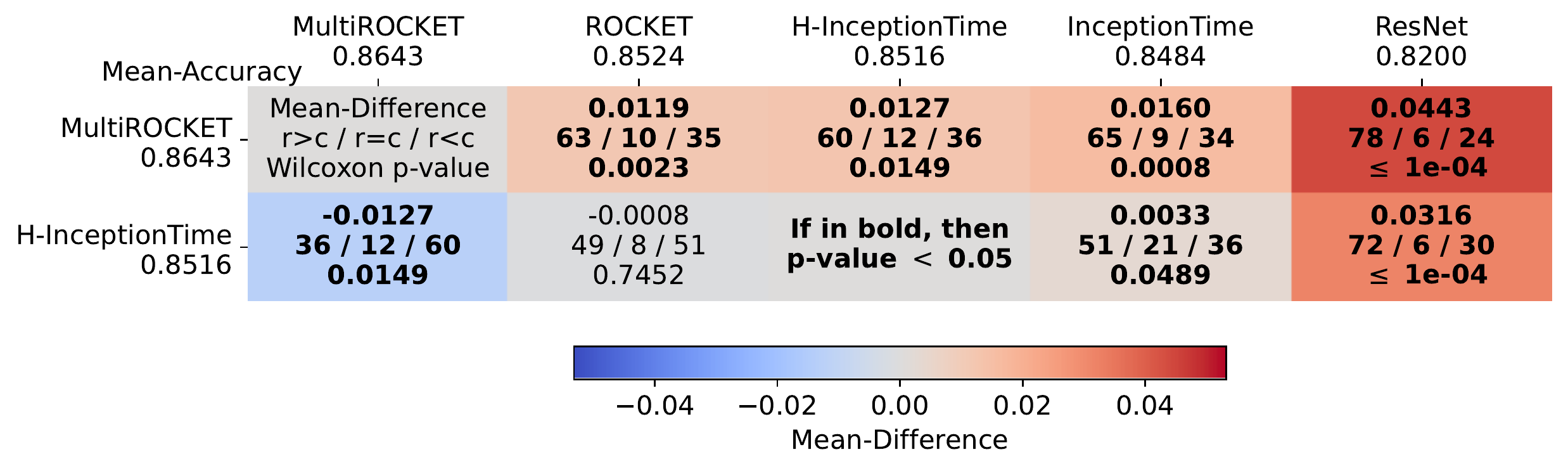}
    \caption{MCM showing pairwise comparisons between MultiROCKET and H-InceptionTime and each of ROCKET, InceptionTime, and ResNet as well as each other on the 108 datasets of the UCR archive. In this setup, the row comparates are a subset of the column comparates.}
    \label{fig:mcm-example-row-in-col}
\end{figure}

An example is shown in Figure~\ref{fig:mcm-example-row-not-in-col}, where the proposed approaches are H-InceptionTime and MultiROCKET, which are compared with other approaches from the literature, InceptionTime, ResNet and ROCKET. In the example shown in Figure~\ref{fig:mcm-example-row-not-in-col}, the proposed comparates are not included in the columns, so the matrix does not show comparisons between the proposed comparates. Alternatively, the proposed comparates can be included in the columns, depending on the user preference (Figure~\ref{fig:mcm-example-row-in-col}).

\section{Conclusion}\label{sec:conclusion}

Measuring progress via benchmarking is ubiquitous in machine learning and computer science.  Analyzing and presenting benchmark results is an important and challenging task.  Existing approaches such as the CD diagram emphasize global (rather than pairwise) comparisons, and are based on mean ranks and statistical significance which, as well as suffering from various well known technical shortcomings, is open to direct manipulation by changes to the comparate set.

We propose a new approach to the presentation of benchmark comparison results for multiple comparates over multiple tasks, the Multiple Comparison Matrix (MCM). MCM emphasizes pairwise comparisons and, by design, ensures that the outcomes for any given pair of comparates are unaffected by other comparates and, therefore, are stable, and not subject to manipulation through changes to the comparate set.

We are not proposing any new statistical method or measure of performance, but rather a change in mindset: moving away from problematic global measures of performance (such as mean rank), and measures of statistical significance which are subject to manipulation, and instead focusing on pairwise comparisons between comparates. We instead order the comparates on the average of the target measure on all the tasks at hand. In consequence, the relative order of any two comparates will not be affected by the removal and addition of other comparates.  We suggest that MCM provides  clear and simple presentation of the results of benchmark comparisons of multiple comparates over multiple tasks while avoiding many shortcomings of exiting approaches. For examples of its use in large benchmark comparisons we refer the interested reader to two recent papers by  \citet{middlehurst2023bake} and \citet{guijorubio2023unsupervised}.

\acks{The work reported in this paper has been supported by the Australian Research Council under grant DP210100072; the ANR TIMES project (grant ANR-17- CE23-0015); and ANR DELEGATION project (grant ANR-21-CE23-0014) of the French Agence Nationale de la Recherche.  The authors would like to thank Professor Eamonn Keogh and all the people who have contributed to the UCR time series classification archive.}

\bibliography{bibliotec.bib}

\end{document}